# Atomically thin gallium layers from solid-melt exfoliation

V. Kochat[1*], A. Samanta[2*], Y. Zhang[1], S. Bhowmick[3], P. Manimunda[3], S. A. S. Asif[3], A. Stender[1], Robert Vajtai[1], A. K. Singh[2#], C. S. Tiwary[1#], P. M. Ajayan[1#]

[1] Materials Science and Nano Engineering, Rice University, Houston, Texas, USA-77005

[2] Materials Research Centre, Indian Institute of Science, Bangalore, Karnataka, India- 560012

[3] Hysitron Inc., Minneapolis, Minnesota 55344, USA

*equal contribution
# corresponding authors

## Abstract

Among the large number of promising two-dimensional (2D) atomic layer crystals, true metallic layers are rare. Through combined theoretical and experimental approaches, we report on the stability and successful exfoliation of atomically thin "gallenene" sheets, having two distinct atomic arrangements along crystallographic twin directions of the parent α-gallium. Utilizing the weak interface between solid and molten phases of gallium, a solid-melt interface exfoliation technique is developed to extract these layers. Phonon dispersion calculations show that gallenene can be stabilized with bulk gallium lattice parameters. The electronic band structure of gallenene shows a combination of partially filled Dirac cone and the non-linear dispersive band near Fermi level suggesting that gallenene should behave as a metallic layer. Furthermore it is observed that strong interaction of gallenene with other 2D semiconductors induces semiconducting to metallic phase transitions in the latter paving the way for using gallenene as interesting metallic contacts in 2D devices.

Corresponding authors: cst.iisc@gmail.com, abhishek@mrc.iisc.ernet.in, ajayan@rice.edu

The isolation of graphene from graphite has fueled research in layered 2D materials as building blocks of 2D nano-electronics (*1,2*). The development of top-down approaches of exfoliation (*2,3,4*) as well as the bottom-up approaches of chemical vapor deposition (*5,6*) and epitaxial growth on specific lattice planes of substrates (*7)* have resulted in the isolation of monolayers of many layered materials such as the transition metal dichalcogenides (*2,3, 8-15*), and other elemental analogues of graphene such as phosphorene (*16*), silicene (*17-21*), germanene (*22-24*), stanene (*25*) and borophene (*26)*. These materials exhibit a wide range of electronic, thermal, mechanical and chemical properties, making them promising candidates for many applications (*27-30)*. The hexagonal crystal structure of the many layered bulk materials, where the layers are held together by the weak van der Waals forces, aids in easy extraction or growth of stable monolayers of such materials. On the other hand, silicene, germanene, borophene and stanene do not have layered bulk counterparts, but yet have been extracted into atomically thin layers regardless of several issues such as stability and scalability. However, unlike graphene, the strong spin-orbit interaction and buckled structure in these materials make them even more interesting as they can have topological insulator phases, which can lead to dissipationless electrical conduction and support quantum spin Hall state (*25, 31-35*). The recent epitaxial growth of single atomic layers of Pb and In on Si (111) substrate has led to the demonstration of superconductivity in the extreme 2D limit (*36-38*). On a similar note, epitaxially grown ultra-thin Fe films have important implications in the development of 2D nano-magnets (*39*). These "metallenes" which are the metallic analogues of graphene can thus form a new class of 2D materials with interesting properties.

Gallium with a rich low temperature phase diagram (*40)*, is a unique metal. Its chemical stability has been used to demonstrate optical phase-change memories with low power consumption (*41-45*). Epitaxially grown Ga films on GaN substrate display superconductivity with increased values of superconducting transition temperature and critical field as compared to bulk-Ga (*46*). These thin Ga films also undergo superconductor-to-insulator quantum phase transition driven by a magnetic field, which also gave the first experimental evidence of Griffiths singularity in a 2D superconducting system (*47*). These suggest that ultrathin Gallium films can offer interesting physics and possible electronic applications. The conventional deposition technique does not provide freedom to select substrate, which is one of the major requirement for multiple applications. Here, we propose, for the first time, a simple exfoliation technique of surface solid layers from the molten phase of Ga which yields atomically thin 2D layers of gallium which we term as "Gallenene" (the monolayer to few layers of Ga atoms). The exfoliation technique from solid-melt interfaces of low melting metallic materials could be a generic technique to obtain atomically thin layers of new materials. In addition to the extraction of gallenene, we also demonstrate using density functional theory, that gallenene can indeed be stabilized on a substrate in two distinct crystallographic orientations, which is in agreement with our experimental observations. The electronic structure of the gallenene films is highly anisotropic with a combination of partially filled Dirac cones and nonlinear dispersive bands which can lead to new fundamental phenomena and a wide range of applications. Gallenene, unlike other 2D vander Waals solids shows strong interaction with substrates as validated by mechanical indentation tests, and band structure and transport properties evaluated by the density functional theory calculations. The substrate-gallenene interaction has been explored in heterostructures of layered 2D materials ($MoS_2$) and gallenene to demonstrate interesting phenomena such as structural phase change in $MoS_2$ and subsequent use of gallenene as contacts.

We first present the theoretical aspects of the gallenene structure. Using density functional theory (DFT) calculation, we first explore the stability of possible layers oriented along specific crystallographic directions in bulk α-Ga, which is found to be the most stable among different bulk Ga phases (*48*). The α-Ga has orthorhombic crystal structure with space group *Cmca* (a=4.58Å, b=7.78Å, c=4.59Å) and contains eight atoms in the conventional unit cell (*48*) (See supplementary Fig. S1). The calculated electron localization function (ELF) on different planes indicates that the first nearest neighbour Ga-Ga bonds are covalently bonded, whereas second, third and fourth nearest neighbour Ga-Ga bonds are more metallic in character. The atoms between two layers of Ga along 100 ($a_{100}$) are bonded by a mixture of metallic and covalent bonds, whereas atoms between two Ga along 010 ($b_{010}$) are bonded by covalent bonds (See supplementary Fig. S2). In order to construct an atomically thin layer of gallenene, a monolayer of Ga along 100 ($a_{100}$) and 010 ($b_{010}$) from α-Ga, were extracted and subsequently relaxed as shown in Fig. 1a-c. The unit cells of the two types of gallenene have four atoms each. After complete relaxation, the gallenene $a_{100}$ transforms to honeycomb structure, whereas gallenene $b_{010}$ retains its structure. The Ga-Ga bond lengths in gallenene $a_{100}$ are not equal as the two nearest neighbors of a Ga atom are at 2.50 Å ($\delta_1$), while, the third one is at 2.51 Å ($\delta_2$). These bond lengths are longer than that of IVB 2D sheets (*49*), due to the relatively larger atomic radius of Ga. Moreover, one of the interior angles of gallenene $a_{100}$ is 122.56°, whereas the other two are 118.9°. This suggests that the hybridization in gallenene $a_{100}$ layer is not completely $sp^2$ inspite of the planar structure. Interestingly, the relaxed gallenene $b_{010}$ structure is found to possess lower symmetry and is similar to zigzag rhombic lattice. The bond lengths are 2.69 and 2.74 Å whereas the angles are 122.24° and 117.59°. Unlike planar gallenene $a_{100}$, the $b_{010}$ forms a quasi 2D multi-decker structure, where two Ga dimers are separated by 1.32 Å along the vertical direction (also defined as buckling height). This buckling is often found in several other 2D materials including penta-graphene, phosphorene, silicene, germanene and stanene (*49-51*). The formation energies (with respect to α-Ga) of gallenene $b_{010}$ (0.36 eV/atom) is lower than $a_{100}$ (0.64 eV/atom), indicating better stability. Here we observe the possibility of having two distinct sheet structures (52) which are exfoliated from the same parent material.

To investigate the dynamical instability of gallenene sheets, we have calculated the phonon dispersions of $a_{100}$ and $b_{010}$. The phonon dispersions of gallenene $a_{100}$ and $b_{010}$ show imaginary frequency indicating the dynamical instability (See supplementary Fig. S3). On the other hand, the phonon dispersion of ~6% uniformly strained gallenene $a_{100}$ does not show any imaginary frequencies, as shown in Fig. 2a. Eventually, the lattice parameter of this strained gallenene $a_{100}$ is identical to the bulk lattice parameter. This indicates that the multilayers having lattice parameter closer to bulk will have a better stability. The negative frequencies in the phonon dispersion of gallenene $b_{010}$ disappear after an application of only ~2% strain (Fig. 2b). This also results into lattice parameters which are closer to bulk, leading to similar conclusions regarding the better stability of the multi-layered structure. Interestingly, this also suggests another interesting possibility of stabilizing monolayers of gallenene (either $a_{100}$ or $b_{010}$) on substrates having lattice parameters closer to that of α-Ga. Further insight into the relative stability of these sheets can be obtained by analyzing the ELF of $a_{100}$ and $b_{010}$ gallenene as shown in Fig. 2c. As expected, the electron charge density of $a_{100}$ (Fig. 2c (iii)) accumulates at the centre between the bonds, which indicates covalent nature between Ga-Ga. The Ga-Ga bonds in $b_{010}$ gallenene show some directionality, which is indicative of more toward covalent character along the y-direction, at the same time in other directions it shows less directionality compared to the y-direction indicative of metallic nature (as shown in Figs. 2c (i) and (ii)). Due to this mixture of metallic and covalent nature, $b_{010}$ gallenene shows better stability. Furthermore, the total energies of these strained gallenene sheets are only a few meV higher than that of the optimized structures, indicating a very shallow minima extending to a wide range of lattice parameters as shown in Fig. 2d. Therefore, growing gallenene on substrates having different lattice parameters would serve as a switch to control the thickness. Overall, from the synthesis prospective, gallenene can offer one of the richest opportunities to tune the thickness and crystallographic type by engineering the strain or the substrates.

Next, we calculate the electronic properties of stabilized (strained) gallenene and the corresponding orbital projected band structure and density of states which are shown in Fig. 2e-f. The gallenene $a_{100}$ and $b_{010}$ are metallic due to the finite density of states at the Fermi energy. Due to planar structure of gallenene $a_{100}$, the $p_z$ bands remain unhybridized and form Dirac cones, which lie above the Fermi level. This is expected as Ga has one less valence electron than graphene, therefore, it requires one extra electron to fill up the holes in Dirac cone. Unlike, graphene, the other $sp^2$ bands are very close to Fermi energy (some of them cross the Fermi energy), therefore, giving rise to anisotropy in the low energy bands. The band structure of gallenene $b_{010}$ is very different. Due to buckling and non-hexagonal lattice, $p_z$ orbitals hybridize strongly with the in-plane orbitals, providing extra stability to this sheet. Furthermore, the bands are highly dispersive with a large bandwidth, implying lighter charge carriers and higher mobilities. The binding energy increases with the number of layers for both the gallenene sheets (See supplementary Fig. S2 e). A detailed study of the band structures as a function of layer number for both $a_{100}$ and $b_{010}$ gallenene sheets shows multiple Dirac cones emerging with the increasing number of layers, indicating the similar bonding nature as for monolayer gallenene (See supplementary Fig. S4-5).

In order to synthesize 2D sheets of lower symmetry materials such as α-Ga (orthorhombic structure), conventional top-down techniques of exfoliation prove difficult. In the current work, we demonstrate a new and unique method of exfoliation of conducting anisotropic Ga ultra-thin films: *Gallenene*. The strength of a metal decreases with the increase in temperature due to larger thermal vibrations and becomes extremely low close to the melting temperature. During cooling of a liquid metal droplet on a solid substrate, the temperature difference between the liquid and the substrate results in heterogeneous nucleation at the substrate-liquid metal interface. The free energy ($G$) of heterogeneous nucleation is related to the free energy of homogeneous nucleation *(53)* as

$$G_{hetero} = G_{homo} \times f(\theta) \qquad \text{-----(1)}$$

where, $\theta$ is the contact angle between solid and liquid surface. For an interface with a larger contact angle ($70^0$ for Ga on $SiO_2$ as listed in supplementary information Table. T1), the heterogeneous nucleation is preferred over homogeneous nucleation resulting in growth of a surface solid crystalline layer above the liquid metal. Due to large differences in the strengths of metals in solid and liquid state, the force required to separate the surface solid layer from the rest of the liquid melt is expected to be much lower. We term this new technique of exfoliation of the surface solid layer from the underlying liquid metal as "solid-melt exfoliation technique". Gallium with a low melting point ($T_m \sim 29.7^0C$) and good wettability with $SiO_2$ makes a good candidate for solid-melt exfoliation. We demonstrate this new technique using a 5μm flat punch indenter integrated with an in-situ nanomechanical testing system, PI88 SEM PicoIndenter (Hysitron, Inc.) in the FESEM (FEI Company). The liquid Ga droplet on Si/$SiO_2$ substrate was cooled down to room temperature ($30^oC$) inside the SEM chamber under high vacuum ($\sim 10^{-8}$ mbar). The onset of solidification is checked by compressing the top surface using the indenter with controlled loading. The snapshot of the exfoliation process during the tensile test is shown in Fig. 3a. During compression, a large sink-in of the surface and good wetting with diamond indenter were observed as shown in Fig. 3a (i). A tensile testing of Ga was conducted by pressing the indenter on the surface and applying strain in the displacement-controlled mode. We observed a good contact between Ga and diamond flat punch until some strain and then the formation of a void at one corner of the interface (Fig. 3a(ii)). The void grows with further loading (Fig. 3a (iii)) and at the end a thin layer of Ga was separated out from the bulk Ga droplet (Fig. 3a(iv)). An atomistic schematic of this process is shown along with the images. The surface was compressed to ~5μm and then pulled out along the tensile direction of the indenter. The load required for the exfoliation was found to be ~30μN for 6μm displacement as shown in Fig. 3b and requires a stress of 1.5MPa to separate the thin Ga layer which is much lower than the maximum tensile strength of solid α-Ga (40MPa). The energy required for the exfoliation is given by the area under the load-displacement curve for exfoliation and was estimated to be 48pJ.

Following the above approach, we now explore the gallenene exfoliation on large (mm-sized) Si/$SiO_2$ wafers. The schematic of this exfoliation process is shown in Fig. 3c. A Ga droplet is first heated to $50^0C$ on a hot plate to achieve uniform melting of the entire droplet. The temperature is then reduced to $30^0C$, which is slightly above $T_m$ of Ga. At this point, a clean Si/$SiO_2$ wafer is brought into contact with the surface of the Ga droplet and then removed. The lower temperature at the $SiO_2$ – Ga interface results in the solidification of the surface Ga layers, which are then exfoliated onto the $SiO_2$ substrate. The optical image in Fig. 3d reveals a mm-sized thin layer of gallenene around a bulk Ga droplet. The continuous gallenene layer is free of any crack or aggregation. The height profile of this gallenene sheet measured using atomic force microscopy (AFM) reveals the thickness to be ~4 nm from the histogram of the step height measured indicating the film to be ~4-6 layer thick (See supplementary Fig. S7). The formation of few-layer thick gallenene films turns out to be an essential condition for stabilizing these sheets as was also revealed from the phonon dispersion calculations. Furthermore, unlike van der Waals solids, the strong mixture of covalent and metallic bonding which can be explored for different applications.

To investigate the scalability and reproducibility of the solid-melt exfoliation technique, we developed a stamping method as shown in Fig. 3e to demonstrate simultaneous exfoliation on multiple wafers with a success rate varying between 80-90% for each batch (See Supplementary Fig. S6). The pie chart in Fig. 3d shows the distribution of the areal coverage of gallenene films obtained from a set of 30 different samples using this technique. Following this, we investigated the substrate effects on gallenene exfoliation by exfoliating these films on various other substrates such as Si(111), GaN(0001), GaAs(111) and polycrystalline Ni using this stamping technique, whose optical micrographs are shown in Fig. 3e. On an average, we find that the exfoliation on Si and GaN yielded flakes of lower thicknesses when compared to the ones exfoliated on GaAs and Ni. A closer microscopic investigation using AFM as shown in Fig. 3f reveals that the gallenene films on Ni were highly discontinuous when compared to the ones on other substrates inspite of similar average surface roughness for the Ni substrate. This indicates that the solid layer exfoliation highly depends on the interaction between the substrate and Ga. The contact angle test performed using liquid Ga droplet on different substrates (See supplementary Fig. S9) clearly reveals a much larger contact angle for Ni which leads to discontinuous layers due to the low wettability of Ga on Ni. The calculated binding energies and ELF analysis explore the chemical interaction of gallenene with the substrate. The details of the change in geometry of the gallenene on different substrates were analysed by DFT calculations and are discussed in subsequent section. It is important to note that unlike other traditional growth techniques such as MBE, this novel technique of exfoliation allows study of ultrathin Ga films on a wide variety of substrates.

The elemental composition characterizations were performed using SEM equipped with energy-dispersive X-ray spectroscopy (EDS) and X-ray photoelectron spectroscopy (XPS). The SEM image in Fig. 4a shows gallenene with a darker contrast on $SiO_2$ and also consisting of brighter regions. The EDS spectrum acquired on the brighter regions (i) shows significant signal originating from Ga - $L_\alpha$ absorption edge. On the other hand, the EDS spectrum from the darker region (ii) also shows this peak for Ga - $L_\alpha$ even though the signal is very weak due to the extreme thinness of this layer. From the elemental composition data obtained from the EDS, we confirmed that these films are indeed Ga sheets. The crystalline nature of these gallenene films was further confirmed by electron diffraction studies in TEM. The gallenene sheets are transferred onto holey-C coated TEM grids using PMMA assisted transfer technique as shown in the bottom panel in Fig. 4a. The composition map obtained using TEM (inset in Fig. 4b) shows uniform distribution of Ga with negligible oxygen verifying the absence of oxidation of these Ga films. The chemical bonding states of Ga was further investigated using XPS (Fig. 4b), which reveals two intense peaks at binding energies of 1117.0 and 1143.0 eV and correspond to $2p_{3/2}$ and $2p_{1/2}$ of Ga respectively. This also reveals that Ga is in the zero valence state which again confirms the absence of any oxide formation during exfoliation. In order to study the electrical transport in these films, we fabricated Ga devices on Si/SiO$_2$ substrate with Ti/Au (10/100nm) contacts using standard e-beam lithography. The two-probe IV-curves, shown in Fig. 4c, obtained from two such devices, Dev_A and Dev_B (shown in inset) displays linear characteristics with resistance values of 1.7 and 4 k$\Omega$ respectively, indicating ohmic contacts to gallenene. Fig. 4c(ii) shows breakdown characteristics of ultra-thin Ga films which shows that current flow higher than -10µA breaks the film into small particles due to Joule heating as shown in the AFM image in inset.

In order to obtain better insights into the crystal structure and various lattice orientations of these gallenene sheets, a detailed TEM analysis on several of such exfoliated samples was performed using a 200kV JEOL 2011 Cryo-TEM. The contamination originating in PMMA assisted transfer is avoided by direct exfoliation of ultrathin Ga films onto TEM grids. From the statistical analysis of these sheets using selected area electron diffraction (SAED), we observe gallenene films of predominantly two different lattice orientations whose representative images are shown in Fig. 4d-e. The low magnification bright field TEM image in Fig. 4d (i) shows an atomically thin film of Ga. The SAED pattern obtained for this Ga sheet (Fig. 4d (ii)) shows that it originates from the (010) lattice orientation of orthorhombic Ga with *a* and *b* having values of 0.2 and 0.26 nm respectively and $\alpha=90^0$. The ratio of *a/b* ~ 0.8, which indicates possible strain in these sheets giving rise to lattice distortion. The HRTEM image in Fig. 4d (iii) shows the lattice arrangement on which the DFT b$_{010}$ structure of gallenene is superimposed. From the HRTEM image, we obtain the lattice spacings as a=0.27 nm and b=0.28 nm, which are in agreement with the lattice parameters of the equilibrium structure obtained from DFT. In Fig. 4e (i), we show the representative low magnification bright field TEM image from gallenene having a different lattice orientation. The inset shows regions with layer thicknesses in the few-layer to monolayer limit. The SAED pattern obtained for this gallenene sheet (Fig. 4e (ii)) shows two hexagons which are rotated with respect to each other by an angle of $6^0$ which could arise due to slightly misoriented gallenene multilayers. This hexagonal symmetry of the diffraction pattern corresponds to the relaxed hexagonal lattice obtained from the (100) plane of the orthorhombic structure of gallenene. The HRTEM image in Fig. 4e (iii) shows this hexagonal structure of Ga with a lattice spacing of 0.2 nm and resembles the a$_{100}$ structure obtained from DFT, which is superimposed on this image. The above experimental results using solid-melt exfoliation results in gallenene with two distinct structures with (100) and (010) crystallographic orientations, which are in agreement with the structures predicted using DFT calculation. The finite thickness observed in experiments further validates the removal of dynamic instability of monolayer by stacking or straining. Single layer of gallenene can be potentially synthesized on suitable substrates having lattice parameters similar to that of α-Ga. The experimental verification of the electronic band-structure changes as a function of varying thickness of Ga films would require detailed ARPES measurements which would be a subject of future studies.

We have explored the stability of gallenene on the various class of substrates such as metals (Al, Ag and Ni), Si and ceramics ($\alpha$-SiO$_2$ and GaN) substrates (See Supplementary information). We find a strong correlation between the structure of the gallenene and orientation, topology and type of the substrates. The calculated binding energy of gallenene on substrates are shown in Fig. 5a. The gallenene a$_{100}$ structure shows good structural stability on the Al, Ag, Si, GaN, and $\alpha$-SiO$_2$, whereas b$_{010}$ show good structural stability on $\alpha$-SiO$_2$ and Ni. The calculated interaction energy per atom of Ga indicates a strong chemical interaction of the gallenene with substrates, unlike the van der Waals solids (graphene, h-BN, MoS$_2$ etc.). The charge transfer analysis shows a charge transfer from gallenene to Si, SiO$_2$, Ag substrates, whereas Al gives the charge to gallenene (See Supplementary information). To get better insight into the nature of chemical bond, we have carried out electron localization function analysis (ELF) (See supplementary Fig. S10). The electrons are localized along the Ga and substrate bonds for Si and SiO$_2$ substrates, suggesting a mixture of ionic and covalent character. On the other hand, Al and Ag substrates show nearly uniform electron charge distribution indicating more metallic character. These strong interactions lead to enhancement of the stability of gallenene. Depending on the strength of the bonding with substrate (stronger covalent in Si and weaker on metallic metals), determines the thickness and stability of gallenene sheets as observed in experiments. Furthermore, due to this strong interaction, the electronic properties gets severely modified depending upon the substrate (See supplementary Fig. S18). Uniquely gallenene offers an entirely different approach to tune the fundamental electronic properties of the sheet itself by changing the underlying substrates. The unique behaviour of gallenene interaction with substrate is further verified using scratch test using a diamond tip as shown in Fig. 5b(inset). The measured force on different substrate (Si, SiO$_2$, GaAs and Ag) clearly show different loads. The adhesive force for gallenene on Si is highest followed by GaN, SiO$_2$ and least is for Ag. The adhesive force (energy) variation is consistent with the interaction energy of gallenene on different substrates.

Plasmonics with 2D materials has achieved a lot of attention recently following the demonstration of surface plasmon resonance (SPR) in 2D layered materials like graphene *(54)*, in addition to other conventional 2D electron gas systems such as perovskites *(56,57)* and semiconductor heterostructures *(58)*. The 2D materials have quite distinct optical properties as compared to bulk due to dimension-dependent plasmon dispersion which is observed in the gallenene sheets as well. The absorption spectrum in Fig. 5c (top panel) shows the SPR peak shift for the gallenene sheets as compared to the bulk Ga. The ultrathin film nature of the gallenene sheets enables light penetration to the Ga-substrate interface which also enables tuning of SPR by modifying the dielectric environment of the underlying substrate. This is demonstrated by comparing SPR of gallenene exfoliated on two different substrates, SiO$_2$ ($\varepsilon_s$ = 3.9) and GaAs ($\varepsilon_s$ = 12.9), where $\varepsilon_s$ is the static dielectric constant. The SPR peak on GaAs having larger $\varepsilon_s$ is shifted to larger wavelengths as shown in Fig. 5c (bottom panel) which is expected from the plasmon dispersion for thin films.

The strong substrate-gallenene interaction can be explored in devices utilizing artificially stacked heterostructure assemblies comprising of various 2D layered materials. For instance, we have studied the interlayer interactions in a heterostructure of monolayer $MoS_2$ and gallenene. The optical micrograph in Fig. 5d (i) shows a large area gallenene partially covering a region with $MoS_2$. The Raman spectra of $MoS_2$ under Ga (Fig. 5d (ii)) shows difference in the relative intensities of the Raman modes of $MoS_2$ ($E_{2g}^1$ at 383 cm$^{-1}$ and $A_{1g}$ at 404 cm$^{-1}$). The photoluminescence (PL) map in the area enclosed in Fig. 5d (i) and spectra are shown in Fig. 5d (iii), which clearly indicate quenching of PL in the $MoS_2$ regions under Ga. These suggest a semiconducting (2H) to metallic (1T) structural phase transition in the $MoS_2$ regions underneath gallenene. Further confirmation is obtained from the XPS spectra of the sample which also shows presence of such a 1T binding state as shown in Fig. 5d (iv). The Mo $3d_{3/2}$ and $3d_{5/2}$ peaks can be fitted using two-peak fitting whereas the S 2p peak is fitted using four peaks corresponding to the 2H and 1T phases of $MoS_2$ (See supplementary information for more details) *(59,60)* While the XPS peaks of Mo $3d_{3/2}$ and Mo $3d_{5/2}$ have values consistent with the 1T and 2H phases of $MoS_2$ indicating a structural transformation to 1T phase; the S 2p peak has broadened and also shifted to lower binding energy values. This could be the result of the strong interaction between the top layer of S atoms in $MoS_2$ with Ga. DFT calculations also show that 2D gallenene can stabilize the 1T phase of $MoS_2$ (See supplementary Fig. S20) This 2H (semiconductor) to 1T (metallic) transformation of $MoS_2$ is of high interest in the fields of catalysis, sensors and 2D metallic contacts to semiconductors.

In the present work we have also explored the thermal and magnetic properties of 2D gallenene. The thermal conductivity (κ) of gallenene was found to be <1 W/m-K which is lower as compared to other 2D materials such as graphene (2000 to 5000 W/m-K at 300K), silicene (5 W/m-K to 50 W/m-K at 300K) and stanene (zigzag and armchair directions are 10.83 W/m-K and 9.2 W/m-K) *(61-64)* (See supplementary Fig. S21-22). Having very low thermal conductivity together with very high electrical conductivity, makes gallenene very unique among the 2D materials. Hence gallenene can be used as a potential on chip electrical connector also acting as thermal barrier in devices. This is promising avenue for the application of the gallenene and requires further experimental exploration. The spin-polarized calculations of mono-vacancy gallenene shows no magnetic ordering. We also do not find any signature of magnetic ordering on substrates as well (See supplementary Fig. S23-24). Apart from the unique properties, we have also theoretically explored different polymorphs of gallenene inspired by recent findings of borophene *(65-67)* as Ga is located in the same group as B, which shows possibility of variety of 2D structures. To explore the possibility of formation of gallenene in borophene-like 2D structure, we have calculated the energy per atom of gallenene by considering all the lowest energy structure of borophenes *(58-60)* (See supplementary Fig. S25). The borophene-like planar and buckling structures of Ga are 81 meV and 41 meV higher in energy compared to the gallenene $b_{010}$ structure, which indicates that $b_{010}$ structure remains most stable. Having said that these structures are only a few meV lower than $b_{010}$ gallenene, therefore, their existence cannot be ruled out by the energy analysis carried out in the current work. Like B, Ga can also offer a richer palette for exploration of 2D structures, having varying functionalities.

In summary, using a combined theoretical and experimental approach, we demonstrate the stability of gallenene sheets having distinct atomic arrangements oriented along two crystallographic (010) and (100) directions. The one-to-one correspondence between the gallenene structures as observed in TEM and the theoretically predicted structures clearly reveals the formation of stable 2D gallenene sheets. The phonon dispersion calculations show that gallenene sheets can be stabilized with bulk lattice parameters. Our new technique for the extraction of gallenene sheets using the solid-melt exfoliation technique could be further extended to exfoliate other metallenes of low melting pure metals and alloys (supplementary Fig. S26 demonstrates possibility of generating ultrathin Sn 2D sheets). Our easily scalable and simple technique allows us to exfoliate on different substrate which can be used for variety of applications. Unlike other van der Waals solids 2D materials gallenene strongly interacts with substrate which is seen in our experiments and theoretical predictions. Based on this this idea, we are able to utilize gallenene contacts to transform the $MoS_2$ from semiconductor to metallic phase resulting in better 2D contacts for devices. The electronic band structure of gallenene sheets show unique combination of unfilled Dirac cone and highly dispersive band near Fermi level and provides excellent opportunities to explore its applications as 2D metals in interconnects, plasmonics, sensors and contacts.

**Acknowledgement**
The Research was also sponsored by AFOSR (Air Force Office of Scientific Research) under Award No. FA9550-14-1-0268. AS and AKS acknowledge Indo–US Science and Technology Forum (IUSSTF), Govt. of India for provided the Indo–US Research Fellowship. VK acknowledge RCQM/Smalley-Curl Postdoctoral Fellowship in Quantum Materials.

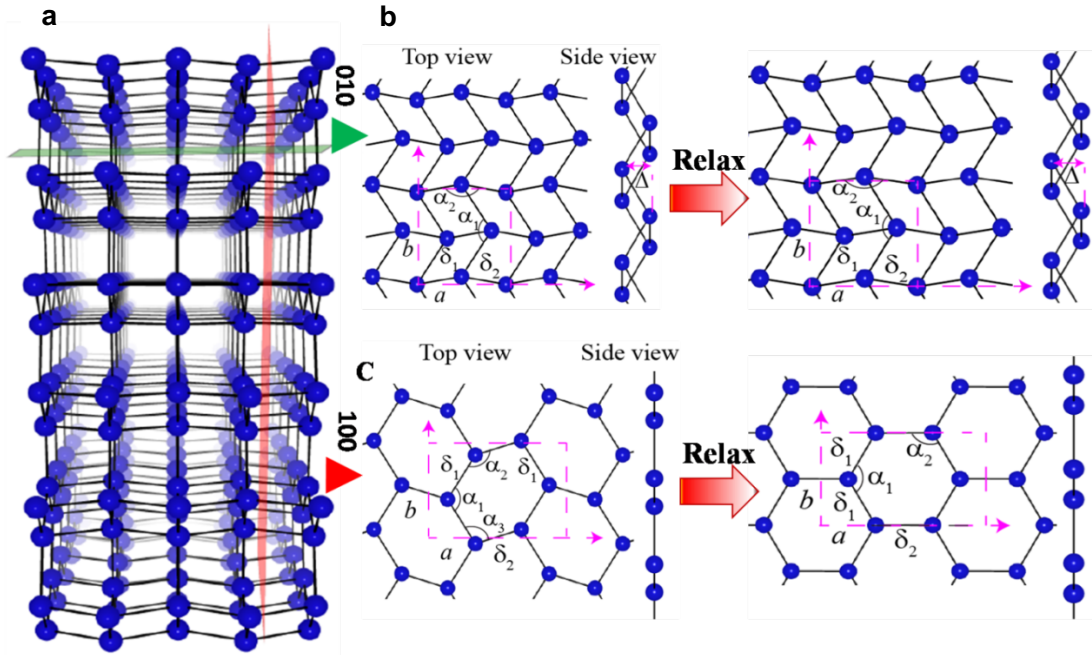

Fig. 1: (a) Crystal structure of α-Ga. (b) Monolayer gallenene structure obtained after cleaving along the (010) direction from the bulk α-Ga as shown by the green plane. On relaxation this forms a distorted rhombic lattice. (c) Monolayer gallenene structure obtained after cleaving along the (100) direction from the bulk α-Ga as shown by the red plane which forms a honeycomb structure after relaxation. The unit cell of monolayer structures is represented by dashed rectangular magenta box. *a* and *b* are the cell parameters of the structure. The bond lengths and bond angles are shown by δ (in Å) and α, respectively. The symbol Δ (in Å) represents the buckling height of monolayer gallenene structure. The cleaved $a_{100}$ structure ($\delta_1$=2.72, $\delta_2$=2.54; $\alpha_1$=139.54$^0$, $\alpha_2$=105.43$^0$, $\alpha_3$=115.03$^0$; Δ=0) transformed to graphene-like structure ($\delta_1$=2.51, $\delta_2$=2.50; $\alpha_1$=122.56$^0$, $\alpha_2 = \alpha_3$=118.9$^0$; Δ=0) after relaxation using density functional theory (DFT). The cleaved $b_{010}$ structure ($\delta_1$=2.77, $\delta_2$=2.72; $\alpha_1$=111.64$^0$, $\alpha_2$=115.03$^0$; Δ=1.46) forms a quasi 2D multi-decker structure ($\delta_1$=2.77, $\delta_2$=2.72; $\alpha_1$=111.64$^0$, $\alpha_2 = $ 115.03$^0$; Δ=1.32) upon relaxation.

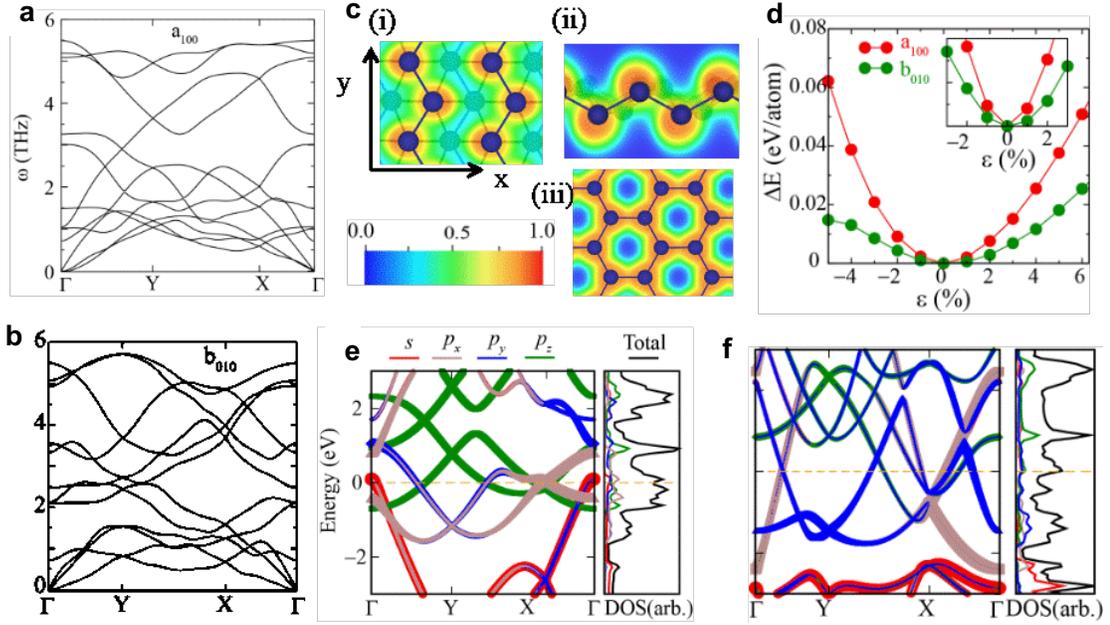

Fig. 2: (a) and (b) are the phonon dispersion of 6% and 2% uniformly strained monolayers of $a_{100}$ and $b_{010}$ respectively. (c) The ELF of 2% strained structure of $b_{010}$ along (i) y-direction and (ii) x-direction and 6% strained structures of $a_{100}$ (iii) monolayers respectively. The ELF is shown by color bar with red (blue) showing maxima (minima). (d) Total energy per atom with respect to lowest energy structure as a function of uniform strain. Inset shows ΔE for a smaller range of strain. (e)-(f) Orbital projected band structure, total density of states (TDOS) and partial density of states (PDOS) of stabilized (strained) $a_{100}$ and $b_{010}$ monolayers with Fermi energy ($E_F$) is set to 0 eV.

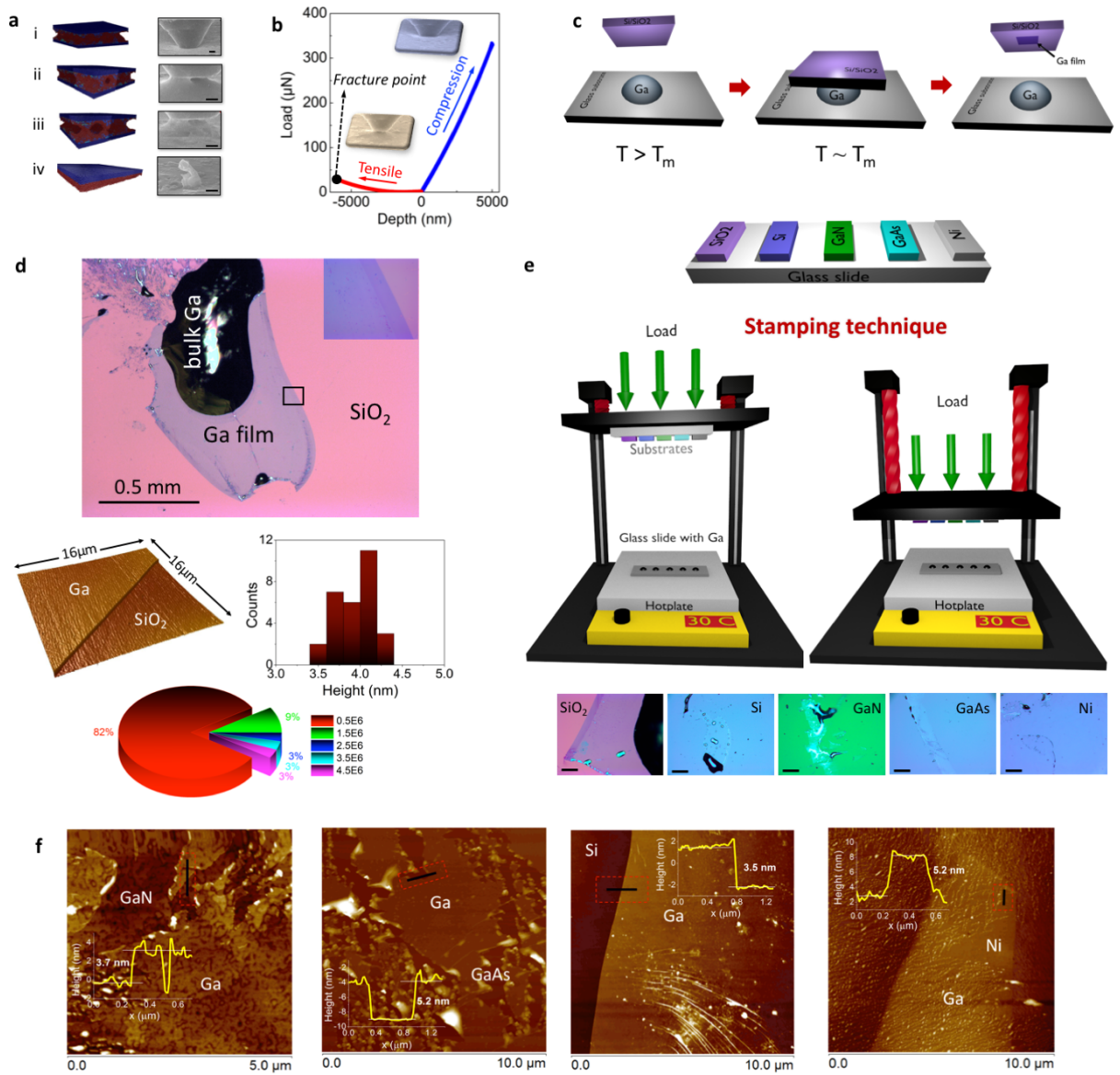

Fig. 3: (a) Snapshots of real-time imaging of gallenene exfoliation using a flat punch indenter inside SEM. The corresponding atomistic schematic of the fracturing is also shown along with this where the red regions correspond to Ga and the blue region is the surface on which Ga ultrathin layer is formed. Scale bar = 1µm (b) Load vs. displacement curve obtained from the in-situ compression and tension test on molten Ga inside SEM. The inset reveals SEM images during tensile and compression of the indenter. (c) Schematic of the proposed solid-melt exfoliation technique of gallenene onto Si/SiO$_2$ wafers. (d) Optical image of Ga sheet on SiO$_2$ wafer showing regions with uniform ultra-thin layers. The AFM measurements on the films reveal the thickness of this film to be ~4 nm as shown from the histogram of step height at the edge along different line scans. The pie chart shows the distribution of the percentage of the area (in µm$^2$) of the exfoliated Ga

films obtained from among 30 different flakes. (e) Schematic of the stamping technique developed for exfoliation on multiple substrates for a fixed load. Optical micrographs of gallenene flakes obtained on various types of substrates is shown at the bottom. Scale bar = 100 μm. (f) AFM images showing exfoliation of gallenene on various substrates like GaN, GaAs, Si and Ni.

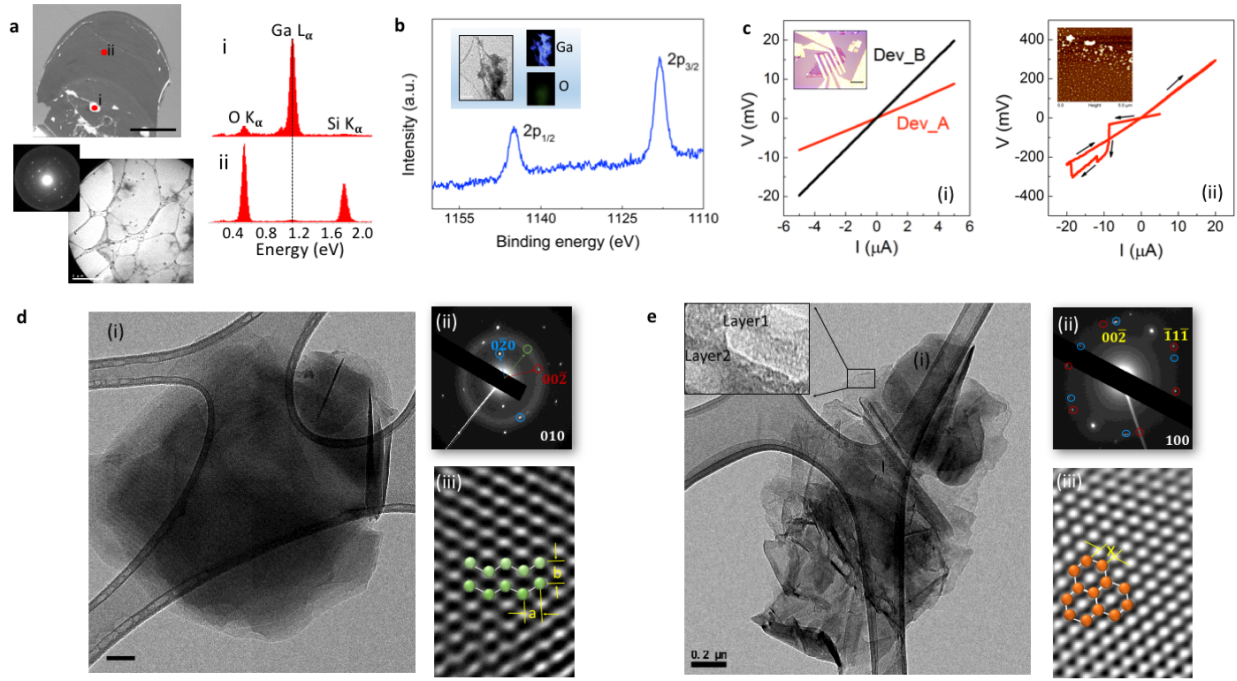

Fig. 4: (a) Top panel: SEM image of the gallenene sheet on SiO$_2$ substrate. Scale bar =50 μm. The EDS spectrum collected from the bulk region (i) shows a strong Ga L peaks whereas the spectrum of the thinner region (ii) also show this peak but with much reduced intensity. Bottom panel: The TEM image of this transferred film along with SAED showing that these Ga film are crystalline. Scale bar = 2μm (b) XPS data showing two intense peaks at binding energies of 1117.0 and 1143.0 eV which correspond to $2p_{3/2}$ and $2p_{1/2}$ states respectively of metallic Ga. Inset: Composition mapping of Ga films in TEM showing presence of Ga with negligible amount of O. Scale bar = 100nm (c) (i) I-V characteristics of two gallenene devices. The inset shows the optical image of Dev_B. Scale bar = 15 μm. (ii) I-V curve showing breakdown of gallenene device at higher currents. The insets show the AFM image of the Ga film after breakdown and its applicability as a 2D electric fuse.(d)-(e) Representative bright field TEM images (i) along with SAED patterns (ii) and HRTEM image (iii) for the gallenene $b_{010}$ and $a_{100}$ sheets respectively. The simulated crystal structure for these two orientations is super-imposed on the HRTEM images. Scale bar = 200 nm.
17

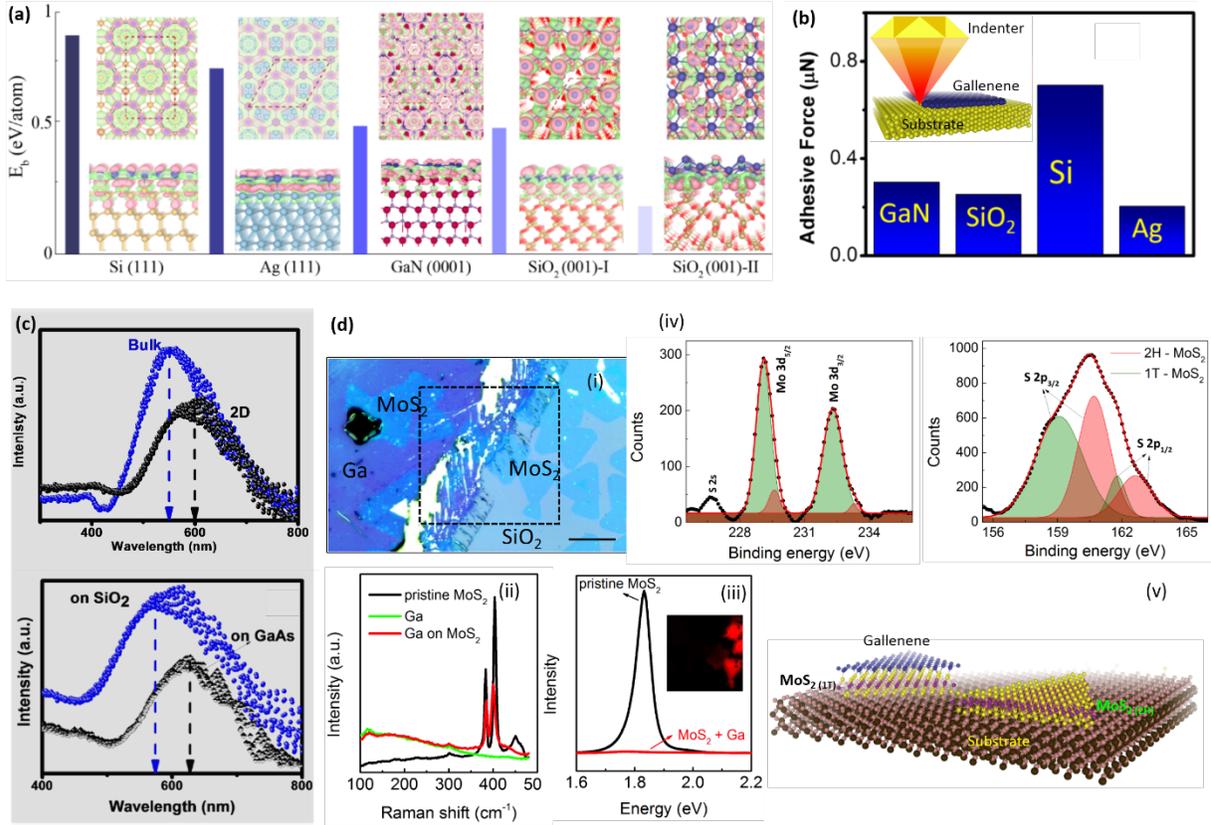

Fig 5: (a) Charge accumulation and depletion of gallenene $a_{100}$ on Si, Ag, GaN, and SiO$_2$ substrates, respectively. Charge accumulation and depletion of gallenene $b_{010}$ on SiO$_2$ substrate is shown in the rightmost panel. The accumulation and depletion of charge are shown by pink and green colors. Top row and bottom row represents top and side view of the structures. The charge uniformly accumulates at the interface of gallenene and metallic substrates, whereas the charge accumulates at the Si (Ga) and Ga bond for Si (GaN) and SiO$_2$ substrates. This implies that the interaction between gallenene and metallic substrate is non-directional towards metallic, whereas the interaction between gallenene and semiconductors (GaN, Si, SiO$_2$) are towards covalent bonding. The bars represent the interlayer binding energy ($E_b$) per Ga atom between gallenene and substrate. The larger value of binding energy indicates greater interaction between gallenene and substrate. Blue, yellow, cyan, brown, purple and red color spheres are Ga, Si, Ag, Ga (in GaN), N and O atoms, respectively. Dotted purple color line represents the unit cell of the structure. The iso-surface value is set at ~ 0.002(e/Å$^3$). (b) The relative interaction (adhesive force) of gallenene with substrate (Si, Ag, SiO$_2$, GaAs) as measured using indenter. (c) The top panel shows the absorption spectra of bulk Ga and ultrathin gallenene sheets. The bottom panel shows the absorption spectra of gallenene sheets on two different substrates, SiO$_2$ and GaAs. (d) (i) Optical micrograph of Gallenene sheet on MoS$_2$. Scale bar = 20 μm. Raman (ii) and PL (iii) spectra of pristine MoS$_2$ and MoS$_2$ underneath gallenene is shown. The PL map of the region indicated by dashed lines in (i) is shown in the inset. (iv) The XPS spectra for the MoS$_2$ region underneath Ga is shown which indicates evolution of 1T-phase of MoS$_2$. (v) Schematic illustrating the idea.